# ELECTRODYNAMICS IN THE ZERO-POINT FIELD: ON THE EQUILIBRIUM SPECTRAL ENERGY DISTRIBUTION AND THE ORIGIN OF INERTIAL MASS

*M. Ibison*

*Institute for advanced Studies at Austin*

*4030 West Braker Lane, Suite 300*

*Austin, TX 78759, USA*

*E-mail: ibison@earthtech.org*

## ABSTRACT

Attempts at an electromagnetic explanation of the inertial mass of charged particles have recently been revived within the framework of Stochastic Electrodynamics, characterized by the adoption of a classical version of the electromagnetic zero-point field (ZPF). Recent claims of progress in that area have to some extent received support from related claims that the classical equilibrium spectrum of charged matter is that of the classically conceived ZPF. The purpose of this note is to suggest that some strong qualifications should accompany these claims. It is pointed out that a classical massless charge cannot acquire mass from nothing as a result of immersion in any EM field, and therefore that the ZPF alone cannot provide a full explanation of inertial mass. Of greater concern, it is observed that the peculiar circumstances under which classical matter is in equilibrium with the ZPF do not concur with observation.

1. ZPF ORIGIN OF INERTIA

1.1    Background

To some degree the ZPF-inertial mass program is a development of work wherein *classical* representations of the ZPF have been used to derive a variety of quantum results. These include van der Waals binding [1,2], the Casimir effect (see for example the calculation and associated comments in [3]), the Davies-Unruh effect [4], the ground-state behavior of the quantum-mechanical harmonic oscillator [5], and, with some qualifications, the blackbody spectrum [1,6-8]. More broadly, besides the hypothesized ZPF-inertial mass connection, these successes have encouraged a 'realistic' classical interpretation of the ZPF [9-13], though the original idea of a classical electromagnetic ZPF seems to date back to Nernst [14]. The focus of this document is exclusively on the electromagnetic zero-point field; the possibility that zero-point conditions of other fields may have a role is admitted but those fields are not considered here.

1.2    Initially massless structureless particles in the ZPF

In its role as the originator of inertial mass, the ZPF has been envisioned as an external, energizing influence for a classical particle whose mass is to be explained. The program has some of the flavor of Mach in that the ZPF provides a 'background' against which the acceleration can be measured. The local properties that ultimately determine the mass of the charged particle enter via the cutoff of the spectral response of the particle to the ZPF [15-17]. These efforts have arisen from within a framework known as Stochastic Electrodynamics [18-20], (SED) wherein a free charged particle is deemed to obey the (relativistic version of the) Braffort-Marshall equation

$$m_0 a^\mu - m_0 \tau_0 \left( \frac{da^\mu}{d\tau} + \frac{a^\lambda a_\lambda}{c^2} u^\mu \right) = eF^{\mu\lambda} u_\lambda,  \qquad (1)$$

(SI units) where $\tau_0 = e^2/6\pi\varepsilon_0 m_0 c^3$, and where $F$ is the field tensor of the ZPF interpreted classically (see [5] for the correspondence between this and the vacuum state of the EM field of QED). It should be emphasized that Eq. (1) is not an ad-hoc extension of classical electrodynamics. If $F$ is the ZPF field tensor operator, then Eq. (1) is a relativistic generalization of the Heisenberg equation of motion for the quantum-mechanical position operator of a free charged particle, properly taking into account the vacuum state of the quantized EM field [21]. As Milonni points out, [22,23], from the standpoint of QED, once coupling to the EM field is 'switched on' and radiation reaction admitted, the action of the vacuum field is not an optional extra, but a necessary component of the fluctuation-dissipation relation between atom and field.

In these ZPF-inertial mass studies, the electrodynamics of the charge in its pre-mass condition has not received attention, presumably on the grounds that the ZPF-energization will 'quickly' render the particle massive, so that the 'intermediate' state of masslessness is of no import. However, letting $m_0 \to 0$ in Eq. (1), one sees that $F^{\mu\lambda} u_\lambda \to 0$, which demands that $\mathbf{E}.\mathbf{v} \to 0$: the massless particle moves always orthogonal to the electric field [24]. It is concluded that if a charge is initially massless, there is no means by which it can acquire inertial mass-energy from an EM field, including the ZPF. This may be regarded as a result of the absence of internal degrees of freedom, whence there is nowhere for the charge to put the energy. Therefore an independent, structureless, massless, point charge *cannot* acquire mass as a result of immersion in the ZPF, and so one must discount the possibility that the given ZPF (given in the sense of a pre-specified field) can alone explain inertial mass of such a particle. This is not to deny that mass may yet emerge from a process involving the ZPF or some other EM field – but it cannot be the whole story.

In recent works by Rueda and Haisch it is argued that inertia – specifically Newton's 2nd Law - arises from scattering of the ZPF by a charged particle [15-17]. Their presentation is particularly attractive because of its simplicity; it appears to make no assumptions about hidden structure or hidden dipoles energized by the ZPF. Concerning the latter possibility, in [16] the authors claim to have derived inertia via an approach "which avoids the ad-hoc particle-field interaction model (Planck oscillator)." (See [25] for a discussion of this later approach of Rueda and Haisch with Dobyns contrasted with their earlier 'parton' model discussed below.) However, consistent with the arguments given above, the charged particles in question are not entirely free of structure: In those works by Rueda and Haisch, an internal structure is implied by their use of a mass-specific frequency-dependent coupling constant between the charged particles and the ZPF. Hence their approach and conclusions do not challenge - but remain consistent with - the assertion above that the classical structureless charged particle cannot acquire mass solely as a result of immersion in the ZPF.

### 1.3 Inertial-mass versus mass-energy

Especially in [25], the distinction has been made between inertia as a 'reaction force' and inertial mass as energy. Hitherto, in this document, there has been no explicit distinction of these two aspects of 'mass', and so some comment seems called for. Specifically, the reader may wonder if the arguments of the previous section apply equally to both aspects of mass, or perhaps they apply only to mass-energy. That is, could it be that the ZPF is the cause of resistance to acceleration, without it having to be the cause of mass-energy? To address this issue: one may observe that the geometric form of the mass-action

$$I = -m_0 c^2 \int \gamma^{-1} dt \qquad (2)$$

simultaneously gives both the mass-acceleration $f^\mu = m_0 a^\mu$ - i.e., the traditional 4-force - contribution to Euler-Lagrange equations, *and* the Noether quantity conserved under time translations $E = \gamma m_0 c^2 = m(v)c^2$ - i.e., the traditional mechanical energy. (Invariance under space translations gives / defines the mechanical momentum.) From the standpoint of *current* physics then, the distinction between the two 'qualities' of mass appears to be one of epistemology: the coefficient of resistance to acceleration due to an applied force *is* - in units where $c = 1$ - the mass-energy.

It is not being claimed that a *physical* distinction between inertia as a 'reaction force' and inertial mass as energy is forever an impossibility. But it is claimed that if it exists, a physical distinction must appear at a level prior to 'extremization' of the geometric mass-action. Whatever the outcome of that debate, of relevance to this document is that, if extremization of Eq. (2) is accepted, the arguments of the previous section apply equally well to the quality of resistance to acceleration as to mass-energy. Specifically: a lone bare charged particle initially exhibiting neither cannot acquire either as a result of immersion in the ZPF.

### 1.4  ZPF-originated inertia of particles with structure

The ZPF has also been envisioned as an external, energizing influence for an explicitly declared local *internal* degree of freedom, intrinsic to the charged particle whose mass is to be explained. In the work by Haisch, Rueda, and Puthoff [26], the charged particle is deemed equipped with some kind of internal, oscillatory, degree of freedom (the 'parton' model). Upon immersion in the ZPF, this ('Planck') oscillator is energized, and the energy so acquired is some or all of its observed inertial mass. Such a particle is not a structureless point in the usual classical sense, and so does not suffer from an inability to acquire mass from the ZPF, provided the proposed components (sub-electron charges) already carry inertia - at least as an assembled unit. If this were not the case – if the assembled components carried no inertia without the ZPF - then the slightest EM influence experienced by dipole would tear it apart, and situation would be as above, wherein the massless parts are unable to benefit energetically from the ZPF. Since mass must already be present before the localized degree of freedom can be activated, the ZPF cannot be the sole originator of inertia, though it could conceivably boost an already present, tiny, localized, mass-energy to the observed value. However, a program such as this would begin to look like an SED version of QED mass-renormalization.

## 2. ON THE CLASSICAL EQUILIBRIUM SPECTRUM OF ELECTROMAGNETISM

### 2.1 Equilibrium between matter and a classical ZPF

The relativistic version of the Braffort-Marshall equation (Eq. (1)) is form-invariant. Additionally, because the ZPF force tensor is statistically indistinguishable in every inertial frame, the *solutions* of the Braffort-Marshall equation are statistically frame-invariant. It follows that the statistics of the Lorentz-Dirac charged particle of SED obeying that equation - including the statistics of the secondary radiation - will be independent of the particle's inertial frame, and therefore SED predicts that the radiation from massive classical charged particles and the ZPF will be mutually self-consistent. However, though permitted mathematically, this possibility is in conflict with observation: an equilibrium configuration between matter and the ZPF viewed from a moving frame leaves the ZPF spectrum (statistically) unchanged, implying the *new* (Lorentz-transformed) matter distribution must be such as to maintain the equilibrium. Since this must be true for any boost, the matter distribution must also be Lorentz-invariant. However, we observe that we do not live in a universe where the matter is distributed in a Lorentz-invariant fashion; on the contrary, the velocity distribution of matter is such as to provide a locally identifiable absolute cosmic reference frame (i.e. via the 3 K cosmic background radiation). It is concluded that the matter we observe is not in equilibrium with the ZPF, classically interpreted, posing a serious problem for SED.

The velocity-independence of the (statistics of the) classical ZPF radiation field permits the immediate deduction that the equilibrium distribution of sources must also be velocity-independent, independent of the details of the interaction between radiation and matter. This relation may be contrasted with traditional classical equilibrium existing between the Rayleigh-Jeans (RJ) radiation spectrum, $\rho \propto \omega^2 T$, and the Maxwell-Boltzmann (MB) matter distribution [27], the analysis leading to which takes place in a single static frame. From a moving frame the same situation would look quite different: the EM spectrum would no longer be an RJ spectrum, and the MB distribution would acquire a net drift / offset velocity.

## 2.2   Support from related work

Boyer has published a derivation of the (QM) blackbody spectrum from the Stochastic Electrodynamics of massive particles *assuming* the existence of a classical zero-point field with its attendant $\omega^3$ spectrum [1,8]. The latter assumption is justified therein simply on the basis of Lorentz invariance of the EM radiation rather than equilibrium between radiation and matter. Boyer's claim does not contradict the conclusions reached above because he does not claim electromagnetic self-consistency. Subsequently, Boyer examined the exchange between a non-relativistic nonlinear dipole oscillator and a background EM field to see if the $\omega^3$ spectrum of the latter can in fact be derived on the basis of classical equilibrium [28]. (In an expanding universe, equilibrium between a classically conceived ZPF and massive charges must be a fossil - from a ZPF-radiation-dominated era!) Boyer's conclusion concurs with the accepted view that the Rayleigh-Jeans and not the ZPF spectrum is the classical equilibrium spectrum [27], supporting the conclusions of the discussion above. His work was repeated and the conclusion confirmed by Pesquera and Claverie [29], and again by Blanco, Pesquera, and Santos [30,31]. (Blanco and Pesquera subsequently drew attention to the interesting fact that analyses demonstrating equilibrium between a Rayleigh-Jeans radiation spectrum and a Maxwell-Boltzmann matter distribution require a cutoff in the Rayleigh-Jeans spectrum [32]. However, these efforts are all non-relativistic; it is possible that the problem will disappear if analyzed relativistically.)

### 2.3 Linear material response to a classical ZPF

The claim that the distribution of charges must be Lorentz-Invariant if they are to be in equilibrium with the classical ZPF should be qualified: Observing that the frequency spectrum of any EM distribution is not modified by a spatially stationary (i.e. statistically static) distribution of dipoles having a perfectly linear response to the in-fields, it follows that *any* frequency distribution of radiation can come to an equilibrium with such dipoles. (A sufficient condition that the radiation momentum distribution also be in equilibrium is that the radiation is isotropic and the matter homogeneous.) On these lines Puthoff has considered massive charges as possible sources of the ZPF, since, in the non-relativistic limit, they have this property of linearity [33,34]. Similarly, Cole has considered massive non-relativistic dipoles for the same job [6,7]. Practically though, such arrangements are unstable, since non-linearities will be present to some degree. Therefore, the radiation spectrum of real massive charges will not reproduce the ZPF: considered classically, field and charges cannot be in mutual equilibrium.

### 3. SUMMARY

Claims that inertia originates from the ZPF are in need of qualification: Inertia cannot originate solely from the ZPF. Schemes in which the ZPF boosts a bare mass to an observed value are not discounted, but should distinguish themselves from the existing technique of mass-renormalization. Claims of spectral self-consistency between charged matter and a classically conceived ZPF are at variance with observation.

**ACKNOWLEDGEMENTS**

The author is very grateful to H. E. Puthoff and A. Rueda for interesting and enjoyable discussions that stimulated much of this work. The author is indebted to the referees for their insightful comments for pointing out errors in the original manuscript.